\documentclass[prl,aps,twocolumn,superscriptaddress,showpacs]{revtex4}
\usepackage{graphicx}

\begin{document}

{\noindent \large \bf Comment on ``Photoluminescence Ring Formation
in Coupled Quantum Wells: Excitonic Versus Ambipolar Diffusion''}

\vskip 3mm

In a recent Letter \cite{Stern08}, an inner ring \cite{Butov02} in
the photoluminescence (PL) pattern from GaAs/AlGaAs coupled
quantum wells is attributed to the Mott transition in a system of
initially photoexcited electron-hole ($e$-$h$) pairs and secondary
excitons. In contrast, in Ref.\,\cite{Ivanov06} the inner ring has
been explained and modelled in terms of in-plane transport and
thermalization of indirect excitons. In our Comment we show that
(i) Stern {\it et al.} considerably overestimate the density of
photoexcited carriers, (ii) the density $n$ of secondary indirect
excitons absolutely dominates over the density of free $e$-$h$
pairs, $n_{\rm e,h} = n_{\rm e} = n_{\rm h}$, and (iii) no Mott
transition occurs in the system.

Equation (2) for the factor $f$, used in Ref.\,\cite{Stern08} in
order to improve the mean field approximation for exciton-exciton
interaction $E_{\rm int}$ (see Eq.\,(1) in Ref.\,\cite{Stern08})
and to evaluate $n$ by the energy shift of the exciton PL line,
cannot be applied for densities relevant to
\cite{Stern08,Butov02,Ivanov06}. The authors completely disregard
the {\it screening} \cite{Ivanov02} of the exciton potential $U =
U_{\rm dd}(r)$ they analyze. The correct expression for $E_{\rm
int}$ is given by $E_{\rm int} = \int n_{\ell}(r) U(r)d{\bf r}$,
where the local density $n_{\ell} = n_{\ell}(r;n,T)$ of excitons
in quasi-equillibrium~\cite{SM} is
\begin{equation}
1 - e^{-T_0/T} = \big(1 - e^{-T^{(0)}_0/T}\big) e^{-(U + u_0
n_{\ell} - u_0n)/(k_{\rm B} T)}
\end{equation}
with the degeneracy temperatures $k_{\rm B}T_0=\pi \hbar^2
n_{\ell}/(2M_{\rm x})$ and $k_{\rm B} T_0^{(0)} = \pi \hbar^2
n/(2M_{\rm x})$, and $u_0 =e^2 d/\varepsilon$ ($d$ is the distance
between the $e$ and $h$ layers). In the classical limit $T \gg
T_0$, Eq.\,(1) reduces to $n_{\ell} = n \exp[-(U + u_0 n_{\ell} -
u_0n)/(k_{\rm B} T)]$. If we now put $u_0 = 0$, the latter
equation results in the low-density approach used in
Refs.\,\cite{Stern08,Zimmermann07}. However, the density-dependent
screening effect cannot be neglected for $n \gtrsim 10^{10}\,{\rm
cm}^{-2}$: It strongly weakens and flattens the input, bare
mid-range potential $U$ at $r \gtrsim d$. In Fig.\,1 we compare
the factor $f = f_{\rm dd}$ used by Stern {\it et al.} for
indirect excitons approximated by classical dipoles (as a remark,
in Eq.\,(2) of Ref.\,\cite{Stern08} the factor $(4 \pi)^{1/3}$ is
missing that results in the strong underestimation of $f$) with
that calculated with Eq.\,(1). We also show $f = f_{\rm xx}(n)$,
evaluated with screening, for a more realistic input potential $U
= U_{\rm xx}(r)$ [inset (a)], taking into account a
quantum-mechanical interaction of undeformed excitons. To
visualize the screening effect, the local form-factor $g =
n_{\ell}(r)/n$ is also plotted in inset (a) against $r$. We
conclude that Stern {\it et al.} have overestimated the density of
indirect excitons by a factor of $5-10$, dealing in reality with
$n \ll 10^{11}\,{\rm cm}^{-2}$, i.e., well below the Mott
transition.

\begin{figure}[t!]
\begin{center}
\includegraphics*[width=5.99cm,angle=270]{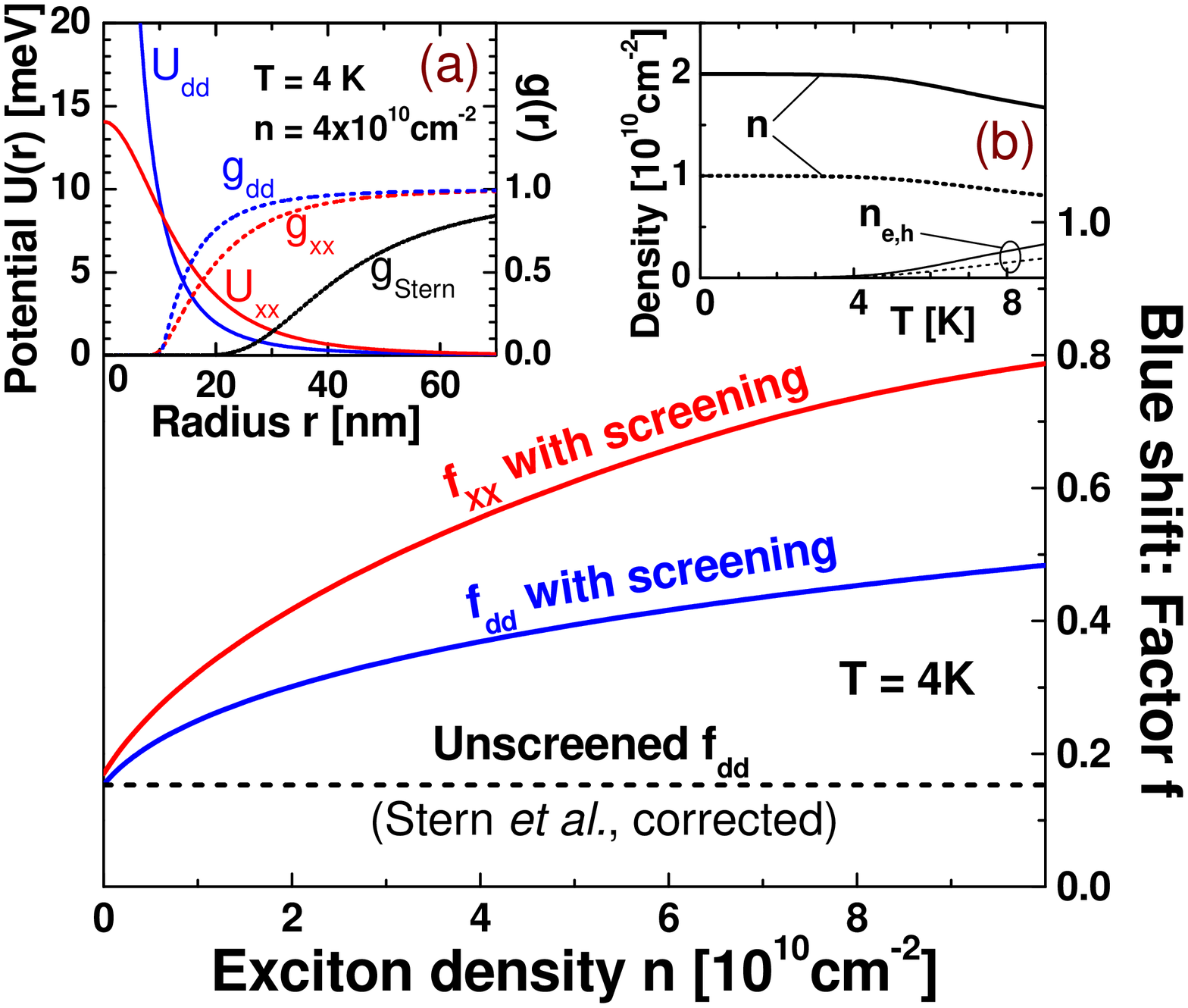}
\caption{The factor $f=E_{\rm int}/(n u_0)$ against the density
$n$ of indirect excitons. Inset (a): The classical dipole-dipole
potential $U = U_{\rm dd}(r)$ and a more realistic $U = U_{\rm
xx}(r)$, along with form-factors $g = g_{\rm dd}(r)$, $g_{\rm
xx}(r)$, and $g_{\rm Stern}(r)$. Inset (b): $n(T)$ and $n_{\rm
e,h}(T)$ for $n + n_{\rm e,h} = 10^{10} \rm{cm}^{-2}$ and $2
\times 10^{10} \rm{cm}^{-2}$, evaluated with the QMAL. To mimic
the experiment \cite{Stern08}, $d = 13.5$\,nm, $M_{\rm x} =
0.215\,m_0$, and $\varepsilon = 12.5$ are used.} \label{Figmass4}
\end{center}
\end{figure}

In inset (b) we show how the total number of $e$-$h$ pairs is
distributed among the bound (exciton) and unbound states,
according to the quantum mass action law (QMAL) and {\em taking
into account screening of excitons by free carriers}~\cite{SM}.
Thus for an effective $e$-$h$-exciton temperature $T \lesssim
6$\,K, relevant to the steady-state experiments
\cite{Ivanov06,Stern08,Butov02}, the absolute majority of $e$-$h$
pairs are in the bound state: $n \gg n_{\rm e,h}$. Furthermore,
$n_{\rm e,h} > 10^{11}\,\mbox{cm}^{-2}$ claimed in \cite{Stern08}
corresponds to the PL line width $\sigma \gtrsim 5$\,meV, in
contradiction to $\sigma \lesssim 2$\,meV reported in
\cite{Ivanov06,Butov02} and observed by Stern {\it et al.}
themselves. The measured diamagnetic shift \cite{Stern08} does not
evidence the presence of $e$-$h$ plasma: The density reduction
with $B$ at a large distance from the excitation spot, needed to
explain the results, is attributed by Stern {\it et al.} to the
Lorentz force acting on free carriers. However, a strong
suppression of transport in magnetic fields is also characteristic
for indirect excitons, due to the exciton mass enhancement,
$M_{\rm x}=M_{\rm x}(B)$~\cite{Lozovik02}.
 \vskip1mm

We thank L.\,V. Butov for valuable discussions.
 \vskip3mm

\noindent A.\,L.~Ivanov$^1$, E.\,A.~Muljarov$^1$,
L.~Mouchliadis$^1$, and R.~Zimmermann$^2$

{\noindent \small $^1$ Department of Physics and Astronomy,
Cardiff University, Cardiff CF24 3AA, United Kingdom}

{\noindent \small $^2$ Institut f\"ur Physik der
Humboldt-Universit\"at zu Berlin, Berlin 12489, Germany}

\vskip3mm


\end{document}